# How Complexity Will Transform Enterprise Information Systems

*Paolo Magrassi - March, 2010 – v4.1*
info@magrassi.net

## Abstract

One "problem" with the 21st century world, particularly the economic and business worlds, is the phenomenal and increasing number of interconnections between economic agents (consumers, firms, banks, markets, national economies). This implies that such agents are all interacting and consequently giving raise to enormous degrees of non-linearity, a.k.a. complexity. Complexity often brings with it unexpected phenomena, such as chaos and emerging behaviour, that can become challenges for the survival of economic agents and systems. Developing *econophysics* approaches are beginning to apply, to the "economic web", methods and models that have been used in physics and/or systems theory to tackle non-linear domains. The paper gives an account of the research in progress in this field and shows its implications for enteprise information systems, anticipating the emergence of software that will allow to reflect the complexity of the business world, as holistic risk management becomes a mandate for financial institutions and business organizations.

## Introduction

The word "complexity" can take many meanings, both in common parlance and in scientific or technological jargons. However, increasingly the most intriguing meaning is the one stemming from non-linearity[1].

A problem is linear if it can be broken into a sum of mutually independent sub-problems. When, to the contrary, the various components/aspects of a problem interact with each other so as to render impossible their separation for solving the problem step by step or in blocks, then the situation is non-linear.

Another way to express the same concept (Feynman 1964) is to use the systems theory definition: a system is linear if it responds with direct proportionality to inputs. This is a system that obeys the superposition principle: the response at a given place and time caused by two or more stimuli is the sum of the responses which would have been caused by each stimulus individually.

**Linearity**

The "systems" and the "problems" that are encountered in nature are essentially non-linear. However, to simplify the studies or for application purposes, one often resorts to linearity as a first-order approximation: if the effects of non-linearity can be considered negligible, a mathematical model can be built that represents the system as if it were linear.

This approach is fecund in many situations. As an example: an audio amplifier is intrinsically non-linear but, within certain frequency limits, it will behave in a linear fashion and be useful for hi-fi; hence, its description throughout audio and hi-fi literature will always be that of a linear system, even if in principle it is not.

Linear models are useful because subject to the hypothesis of linearity many natural systems resemble one another: their behaviour can be described with the same equations even if the contexts are very different, such as mechanics, electronics, chemistry, biology, economics, and so on. A linear oscillator is a model described by the same mathematical equation, whether it be a metal spring, an electric circuit or a standalone El Niño. (Complex systems, on the contrary, each have their own mathematical formalization and, in many cases, not even that: equations are substituted by numerical computer simulations.)

---

[1] We are using the dynamic/systemic view of complexity. Another one is possible, i.e. the computational/structural one. Ultimately related to Gödel incompleteness theorems, the structural view is predominantly adopted in information theory and computer science, where it has to do with the computability of algorithms. The two views are ultimately connected, although in subtle and convoluted ways, via the concept of entropy.

**Non-linearity**

Gigantic scientific and technological advances have been made using simplifying linearity assumptions, before computers started allowing to venture into non-linear territory. This is how "complexity science", a.k.a. "complexity theory", was born.

There had been, in fact, several explorations of non-linear territory made by scholars since the 19th century. As an example, French mathematician and physicist Henri Poincaré was the first (Poincaré 1890) to discover and describe how an apparently simple system subject to deterministic laws, such as that composed of three orbiting celestial bodies (e.g., Sun, Earth and Moon), can exhibit a complex (chaotic) behaviour. Other scholars, including, e.g., Alexander Bogdanov, Norbert Wiener and Warren Weaver, made advances and contributed creating complex system thinking in the first half of the 20th century (Weaver 1948).

However, the field has acquired new lymph only with the advent of electronic computers, as they allow to simulate whenever mathematics does not do the job because equations are unknown.

Fundamental, in this respect, was the work of mathematician and climatologist Edward Lorenz. Lorenz made apparent (Lorenz 1963) and formalized the problem that Poincaré touched upon in his three-body system: When observing the evolution of a complex system (i.e., its trajectory in state space), *finite* variations may originate from *infinitesimal* variations in the initial conditions.

In other words, even two infinitely similar beginnings will look different in the future, because the evolution of the system will differ substantially in the two cases, with a divergence ever-larger in time. If we were to model the evolution of continental weather, it may make a difference, to the effect of a probability of a tornado in Texas, whether or not «a butterfly flaps its wings in Brazil». Making long-term forecasts becomes impossible.

**Examples of complexity**

A striking exemplification of the above is to be found in another small system such as the one composed of a population of predator animals, a population of preys and the food available to the latter. A linear model turns out simplistic and inadequate for the situation: the population of preys is a function of the predators' population but, in turn, the latter will expand and contract based on the availability of preys. These, on the other hand, depend of the availability of food, and eating too much of it would cause the preys population to contract, possibly beyond sustainability.

The "preys – predators – food" system is intrinsically non-linear: none of its components may be studied in isolation from the others. And indeed, the Lotka-Volterra equations are a classical example of simple non-linear model of an ecological situation. Purposely simplified, the Lotka-Volterra model may lead to the formulation of the so-called logistic map (May 1976):

$$x_{n+1} = r\, x_n\, (1 - x_n)$$

with $x_0$ representing the initial condition, i.e. the initial population at discrete time interval $t_0$.

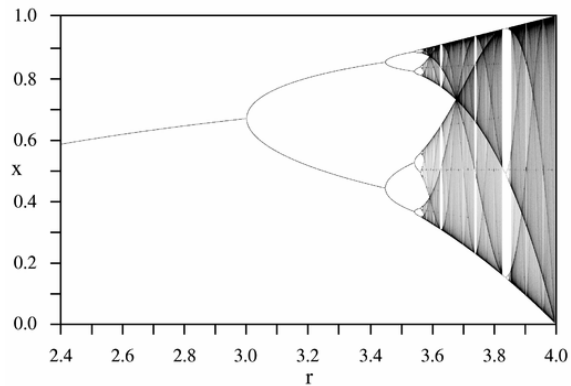

**Figure 1: The logistic map**

By varying the parameter *r*, a number of weird things happen (see Figure 1), particularly for *r* equal or greater than 3.57, when periodic oscillations (corresponding to the increases and decreases of a population depending on ecological conditions) start being replaced by pure chaos.

Thus, a "toy" ecological problem can, like Poincaré's three bodies, end in chaos. In addition to sensitivity to the initial conditions, a second (and related) property of complex systems is indeed *deterministic chaos*: the underlying laws (physical, biological, etc.) may be orderly and even deterministic, yet chaotic behaviour is possible.

**Complex systems can surprise**

A third essential property of complexity is *emerging behaviour*: even when the laws governing its components are well-known, a complex system may show a behaviour that cannot be explained on those grounds.

Popular literature on complexity tends to furnish us with examples from the living world or other high-level natural systems (flocks of birds and colonies of ants behave in ways inexplicable based on what we know of the capabilities of the individuals), however emerging behaviour has been known to physicists since Phil Anderson showed it in the case of groups of electrons in a semiconductor (Anderson 1972).

This observation, along with the chaotic behaviour seen in very small deterministic systems, shows how intimately the natural world is permeated with complexity. It also warns us that knowing "fundamental" laws (such as those concerning elementary particles or individual economic agents) is insufficient to understand the systemic behaviour of interacting components, because in non-linear systems a new behaviour may emerge. We need to know both the fundamental laws and the systemic ones.

**The origins of complexity**

What makes a system complex are the interactions between the components –the ultimate cause of non-linearity (Bridgman 1927)–, not the number of components themselves.

A system made of many non-interacting parts is not complex (from Latin *complector*: to encircle, to embrace firmly, to comprise, to unite under a single thought and a single denomination) but merely complicated (from *complico*, to fold): it takes a long time to unfold it, to solve it, but it can be done step by step.

A complex system/problem, instead, is hard to tackle because, in addition to the laws governing the components, we need to study the system's overall behaviour: the analytic approach must be complemented with the holistic one.

Clearly, though, an increasing number of interacting components will give raise to increasing complexity, possibly exponentially: complex *and* complicated makes things worse.

# Complexity and economics

This is why, in today's economic world, we are more and more often confronted with complexity: it is caused by the number of interactions within (and between) the systems that surround us (Lo 2009), which are complex and complicated at the same time. Sources of complexity include:
- global financial systems;
- networks (such as the Internet, power grids or transport networks);
- enteprises, as these are increasingly interconnected in supply and demand chains, ecosystems, and "clouds";
- consumers, since their behaviours are mutually influencing due to connections of all sorts (TV, mobile communications, Web, social networks, email).

**Linearity dominates the current economic paradigm**

This complexity compounds with mounting debate around the dominant paradigm in economics, the Rational Expectations Hypothesis (REH), which assumes that economic agents (workers, consumers, firms, etc.) behave rationally and take all available information into account in forming expectations of economic events (such as the price of a stock). And firmly on top of the REH rests the other dominating assumption, i.e. that markets are efficient, cannot be "fooled", and smooth out all imperfections in the large numbers (Efficient Markets Hypothesis, EMH).

This paradigm is challenged on at least two fronts. One is behavioural finance/economics (Kahneman 1979), where researchers have proved that *homo economicus* is far from making rational choices, much less to have rational expectations of future events. The second front is indeed the complexity-driven camp.

The REH assumes that any market is always in the surroundings of its equilibrium (a point, line, surface, volume or hypervolume in a state space), and that perturbations around such state may only be small and linear. The perturbations are small because the actual price of a good (commodity, stock, …) is at any time given by

$$P = P^* + \varepsilon$$

where $P^*$=price expected and $\varepsilon$ is an infinitely small quantity. They are linear because $\varepsilon$ is independent of $P^*$ and, more importantly, because the resulting statistical models assume that the risk of every individual investment/asset can be assessed in separation of the risk of other investments/assets: each individual risk is a Gaussian distribution (of the values of the variable/asset under consideration) where the standard deviation is the measure of the asset's volatility.

**Need of an holistic approach**

It follows that REH may be good at assessing the risk of each and every *individual* investment, but the assumption of zero interdependence between investments is an over-simplification that depicts an inherently complex domain as if it were a linear one, thereby causing statistical models to underestimate –or miss altogether– systemic effects.

It can be argued (Bouchaud 2008) that markets of the 21st century global world can no longer be modelled as linear systems. It is therefore not a surprise that REH-based economics be clueless at anticipating shocks such as the one in Asia 1997 or the subprime crisis culminated in the 2008 financial meltdown. (REH/EMH proponents counter-object that the 2008 financial crash was not but the efficient-market anticipation of a to-be economic recession. According to this view, finance was the victim, not the cause, of the economy meltdown (Cassidy 2010)).

The non-linearity of the global financial system cannot be captured by most REH constructs, and in the increasingly interconnected "economic web" (Kauffman 1995) it is essential to have estimates of systemic, global risk, not just the many views of all individual variables: this is because chaos may originate from the combined actions of different agents through their interconnections. Emerging behaviour may take place, it should be noted, whether or not the individual economic agents are believed to be rational: all it takes is that they interact and influence each others significantly.

It follows that an holistic approach to financial markets or economic systems can hardly be based on REH/EMH and that a radical new view of the world may be needed. While economists are of course aware of the limitations of the current paradigm (Scarf 1960) (Sonnenschein 1972) (Stiglitz 1975) (Anderson 1988), proposals for its radical reform seem to come more frequently from outside the economic community, or from its outskirts.

## Econophysics

Radical new views are being attempted in econophysics, a discipline born in the mid-1990's that is trying to a) import more elements of empirical research in economics (a discipline currently resembling more mathematics than physics) and b) have economics research adopt some of the methods devised in the natural sciences for describing *complex* systems. (A broader, if banal, view of econophysics describes it as a discipline aimed at using in general the mathematical methods of physics in the economy: however, this has been done all the time since Leon Walras and Vilfredo Pareto, in the 19th century).

Non-linearity, systems operating far from equilibrium, and "organized disorder" (deterministic chaos, emerging behaviour) are the tools of the trade in econophysics today: just about the opposite of what happens in the REH/EMH paradigm, where markets have no internal dynamics (no interconnections between individual agents and choices) and chaos may only be stochastic, i.e., the "disorganized complexity" (Weaver 1948) of brownian motion and equilibrium thermodynamics. Some physicists and a few "maverick" economists are increasingly recognizing in complex models of the physical world (especially, although not only, condensed matter) situations that resemble economic or financial contexts.

**Proposed new models**

One such model is that of spin glasses (Parisi 1983) (Anderson 1988), a domain characterized by an extreme fragility with respect to small changes in parameters and the de facto absence of equilibrium. Another one, and considered more promising by many (Bouchaud 2009), is heterogeneous mean-field approximation (Mézard 2009), where the problem of N interacting particles is treated as that of a single particle immersed in a chosen field of forces. Yet another model is the one presented in (De Laurentis 2009), where a thermodynamics of non equilibrium approach is taken to show the analogy between Prigogine's dissipative structures and financial markets.

A vast class of models are agent-based models: here, the domain under investigation is explored via computer-based simulation, a technique much more accepted in physics (and chemistry and biology) than in economic research, where, according to many, it should be pursued more actively.

Numerical investigation of a model does not *prove* anything, yet provides a formidable tool, a «telescope of the mind multiplying human powers of analysis and insight just as a telescope does our powers of vision» (Buchanan 2008).

In the case of econophysics studies, simulations often consist in defining economic agents (people, firms, banks, regulators, …) and the rules of the game, then letting the game run to study the possible outcomes. Agent-based simulations, sometimes also referred to as complex adaptive systems or cellular automata, were inaugurated by John H. Conway (Gardner 1970) and first formalized in (Holland 1975). It is impressive to notice how, given extremely simple "organisms" or "agents" and a few governing rules (1 organism and 4 simple rules in Conway's game), rich and complex forms of "life" can evolve over time.

Econophysics simulations of this sort often lead to situations very different from the perpetual quasi-equilibrium of efficient markets. For example, catastrophic meltdowns can take place abruptly, something that in EMH models may happen only with infinitely low probabilities (contrary to empirical evidence). Examples of this kind of simulations are to be found in, e.g., (Thurner 2010), (Westerhoff 2004) or (Macal 2004).

# The enterprise level: the challenge

As the world's complexity increases, crises, including nearly-catastrophic ones such as that of 2008, can be expected to become more and more frequent. Such crises may include sudden perturbations in financial markets or segments of the economy, or shocks in increasingly complex supply chains and industrial ecosystems (Sodhi 2010).

Shocks can propagate in unexpected ways and affect systems or subsystems deemed to be only very loosely coupled with the one first hit. A sudden financial crisis may mean death for just about any financial company and, for that matter, companies of any kind, as we witnessed in 2008-2009 when all OECD countries experienced sharp declines in Gross Domestic Products.

Although the final word rests on the economic research community (econophysicists included), and despite the fact that econophysics results seem a bit overstated nowadays (Gallegati 2006), it seems plausible that there be a need for more-realistic economic laws allowing us to achieve a better grasp of heavily non-linear phenomena which we currently treat as if they were linear and, presumably because of that, seem not to conform to any known statistics.

If such laws are discovered, it will take a long time, due to the intellectual challenges implied, the scarcity of methods for tackling complexity even in the natural sciences, the intricacies of interdisciplinary collaboration, the impossibility of controlled experiments in macroeconomy and the consequent dominance of axiomatic approaches and attitudes in economics, the current limited participation of economists in econophysics research, and, last but not least, the inertia of academia (it is easier to win a tenure or a PhD assignment by participating in a study that resonates with the dominant paradigm than by venturing in distant and subversive territories).

**Systemic risk management and complexity as a proxy of risk**

While we wait for a "new paradigm for economics" to develop, whether from econophysics or out of elsewhere, the increasing uncertainty of their environment will force businesses to implement holistic risk management, i.e. the capability to assess the risk of an entire system as opposed to just that of its components (Lo 2009). This is a must for financial institutions, but a requirement for corporations at large, too, since survival, not growth, may be the name of the game in the future business atmosphere, dominated by uncertainty and fragility (Marczyk 2009).

The "system", in this case, is as simple to name as is difficult to monitor and control: it is the firm itself, in its whole entirety and complexity, immersed in a wider ecosystem.

Even in financial companies, risk management (RM) is traditionally split into separate concerns. Operational RM on one side, with the task of ensuring process integrity, continuity of business and security; and credit and market RM on the other, to protect and exploit counterparty exposures and market developments.

All this has historically been limited to audit, compliance, treasury and insurance areas, dominated by a passive/reactive attitude, with little or no coordination across the enterprise, a lack of analytical tools, and modest information technology (IT) support (Gartner 2008).

As it is understood today, holistic (or "integrated", or "systemic") RM is to become a process that permeates the entire firm and its ecosystem/supply chain, spanning not just risk areas but also strategic goals. From that perspective, there is a need to develop, among other things:

- real-time predictive analysis and reporting of risk, including agent-based simulation;
- infra- and inter-enterprise workflow connectivity and sharing of information;
- common risk management processes through the supply chain/ecosystem;
- XML-based standards for risk data definitions;

and possibly an integrated-risk data warehouse for the enterprise.

# Conclusions: the opportunity

Therefore, the intrusion of complex thinking and holistic/systemic risk management in the enterprise will entail a) the acquisition and/or building of sophisticated software tools and b) the definition of adequate processes and the integration of the new tools with the enterprise's information systems (IS).

Software tools will consist of a new generation of business intelligence products which, we predict, will be based on complexity technology, as complexity is the driver of systemic risk.

Fields like electronics, optronics, avionics, chemistry, biology, ecology and econophysics are providing plenty of examples where non-linear system dynamics models and simulation scenarios are used to control, manage and exploit complex systems. Enterprises will use these "probes" to explore their environment searching for potential complexity peaks, as well as to assess their own (and their supply chains') levels of complexity, fragility and uncertainty.

**Measuring complexity**

Tools will need to

- know ways of communicating sophisticated system concepts, variables and parameters to business people;
- provide quantitative measures/estimates of complexity (and consequent risk) that be simple, synthetic and actionable;
- effect what-if analyses, for example via agent-based simulation, of potentially chaotic situations;
- provide pictorial representations of complexity scenarios (e.g.: depict attractors in state space and the expected probabilities for the system to hang around them);
- analyse the enterprise's complexity drivers, then its partners', then merge them all;
- analyse the entire world's complexity drivers and dynamics, using mash-up applications that leverage various Web sources;
- proactively propose countermeasures aimed at reducing systemic risk.

One major challenge in view of developing this class of tools is to decide how to assess/measure complexity. Early approaches have already materialized in industrial products. As an example, (Ontonix 2008) presents a tool that quantifies complexity of both the enterprise and its environment and provides simple and manageable outputs: the underlying methodology is undisclosed, and personal communications received by this author seem to indicate that the tool assesses complexity by looking at the degree of correlation between up to hundreds of randomly selected parameters of an enterprise's performance.

**An emerging new class of software and market**

In addition to complexity assessment tools, which will be provided by specialized vendors (including, one would presume, the incumbent business intelligence companies), there will be a need to establish an enterprise and an inter-enterprise architecture and integration framework allowing to monitor systemic risk. This means that, in addition to ad-hoc processes, there will be software spanning all or most of the enteprise's IS.

Therefore, the application of complexity-based risk management technologies to enterprises in all industry sectors is a huge opportunity for the software industry. This will be a significant market not just for specialized and niche IT vendors providing new risk management tools, but also for major enterprise software (business intelligence, supply-chain management, etc.) vendors to integrate their software with the former and to provide sustainable human interfaces.

Growing economic complexity constitutes at the same time a challenge and a blessing for IT researchers, practitioners and vendors.

**A bigger picture?**

But the systemic risk management transformation and the growing awareness of the role being played by complexity will probably have an even bigger impact.

The accelerated shrinking of business planning horizons, the increasing financial and economic turbulence, and the impending chaotic behaviour of ecosystems may create a situation where the primary business goal will be *survival*. Keeping overall risk below a certain numeric level could then become as valuable, and as much remunerated, as achieving EBITDA (*earnings before income tax, depreciation and amortization*) targets, because risk is indeed the uncertainty of achieving such goals.

If that happens, new methods of enterprise success and valuation will emerge, and this would transform information systems radically. Today, information systems are optimized for enterprise growth and profit: but if these become secondary objectives and mastery in uncertainty management rises to be the key performance indicator, a Copernican revolution may take place.